\documentclass[
 preprint,
 amsmath,amssymb,
 aps,
]{revtex4-1}

\usepackage{physics}
\usepackage{graphicx}
\usepackage{dcolumn}
\usepackage{bm}

\usepackage{siunitx}
\usepackage{xcolor}
\usepackage{soul} 
\usepackage{subcaption}
\usepackage{graphics}
\graphicspath{{Figures/}}

\usepackage{lipsum}

\begin{document}

\title{Broadband frequency conversion of ultrashort pulses \\ using high-$Q$ metasurface cavities}

\author{Timo Stolt and Mikko J. Huttunen} 
\address{
    Photonics Laboratory, Physics Unit, Tampere University, FI-33014 Tampere, Finland
}

\date{\today}

\begin{abstract}
Frequency conversion of light can be dramatically enhanced using high quality factor ($Q$-factor) cavities. Unfortunately,  the achievable conversion efficiencies and conversion bandwidths are fundamentally limited by the time--bandwidth limit of the cavity, restricting their use in frequency conversion of ultrashort pulses. Here, we propose and numerically demonstrate sum-frequency generation based frequency conversion using a metasurface-based cavity configuration that could overcome this limitation. The proposed experimental configuration takes use of the spatially dispersive responses of periodic metasurfaces supporting collective surface lattice resonances (SLRs), and can be utilized for broadband frequency conversion of ultrashort pulses.
We investigate a plasmonic metasurface, supporting a high-$Q$ SLR ($Q=500$, linewidth of 2 nm) centred near 1000~nm, and demonstrate ${\sim}$1000-fold enhancements of nonlinear signals. Furthermore, we demonstrate broadband frequency conversion with a pump conversion bandwidth reaching 75~nm, a value that greatly surpasses the linewidth of the studied cavity.
Our work opens new avenues to utilize high-$Q$ metasurfaces also for broadband frequency conversion of light. 

\begin{description}
\item[DOI]
\end{description}
\end{abstract}

\pacs{Valid PACS appear here}
                             
\maketitle
 
\noindent 
Since the construction of the first laser in 1960, lasers have been the most common instruments to generate intense, coherent, monochromatic, and directional light~\cite{Maiman1960}.
When applying specific techniques, such as $Q$-switching and mode-locking, lasers can be used to generate ultrashort pulses with extremely high peak intensities and pulse durations down to a few femtoseconds.
A major drawback of ultrashort pulse lasers is that they lack in terms of tunability.
Visible and infrared spectral regions are commonly accessed by utilizing nonlinear frequency conversion resulting from nonlinear processes, such as second-harmonic generation (SHG), sum-frequency generation (SFG), or difference-frequency generation (DFG)~\cite{BoydBook2020}.
Unfortunately, nonlinear processes are, by their nature, extremely inefficient.
Conventional nonlinear optical devices overcome this drawback by utilizing phase-matching techniques and optical cavities~\cite{BoydBook2020}.
Even though these techniques solve the efficiency problem, their operation bandwidths are often quite narrow, restricting their use in frequency conversion of ultrashort pulses with broad spectral features.
This trade-off between conversion efficiency and bandwidth can be solved using adiabatic frequency conversion~\cite{Suchowski2008,Suchowski2011}.
However, such techniques rely on long propagation lengths and complicated phase-matching schemes in the nonlinear medium, motivating to seek for alternative approaches.

Recent progress in the fabrication of nanostructures has enabled the development of a novel material class called metamaterials~\cite{Soukoulis2011}.
They are artificial structures consisting of nanoscale building blocks such as nanoparticles (NPs) and gratings.
Interestingly, the optical properties of metamaterials can be engineered by tuning the properties of the building blocks, such as their size and shape, during the fabrication process.
As a result, metamaterials can exhibit exotic properties, such as negative index of refraction, epsilon-near-zero behavior at optical frequencies, and nanoscale phase-engineering capabilities~\cite{Klein2006, Alu2007, Zhang2009,Genevet2017}. 

Plasmonic metasurfaces consisting of metallic NPs have recently shown the potential for enhancing nonlinear processes in nanoscale structures~\cite{Kauranen2012review}.
Metal NPs exhibit collective oscillations of the conduction electrons
giving rise to localized surface plasmon resonances (LSPRs)~\cite{Maier2007},
which result in an increased local field near the NP, subsequently enhancing the nonlinear response~\cite{Lapine2014, ButetReview2015, Li2017, Rahimi2018, Huttunen2019review}.
However, LSPRs are associated with low quality factors ($Q$-factors, $Q\,{<}\,10$) due to the high ohmic losses associated with plasmon resonances. 
Fortunately, periodically arranged NPs exhibit surface lattice resonances (SLRs)~\cite{Kravets2018,Utyushev2021}, that are associated with narrow spectral features and thus of much higher $Q$-factors ($Q\,{\approx}\,2300$) than LSPRs~\cite{Saad2021}.
Therefore, SLRs can result in dramatic local-field enhancements and consequent enhancement of nonlinear responses~\cite{Michaeli2017,Hooper2018,Huttunen2019}. 

Despite the potential of utilizing SLR-based metasurface cavities for frequency conversion, their behavior is restricted by the time--bandwidth limit associated with optical cavities~\cite{Fan2003}. Further enhancement of the local fields present near the NPs by designing SLRs with record-high $Q$-factors will simultaneously limit their use to frequency conversion of spectrally narrow laser sources~\cite{Saad2021}. Therefore, use of high-$Q$ metasurfaces is seemingly restricted to spectrally narrow laser sources and subsequent nonlinear applications.

In this work, we propose an experimental configuration to achieve broadband frequency conversion with a single plasmonic metasurface supporting a high-$Q$ SLR cavity ($Q\approx500$, center wavelength 1002 nm, linewidth of 2 nm). The proposed setup utilizes a temporal-focusing scheme that first separates an incident broadband laser beam into separate spectral components that interact nonlinearly with the metasurface. The spatial dispersion of SLRs allows us to couple these different spectral components of the incident beam, arriving at the metasurface at different incidence angles, optimally with the SLR of the metasurface. After the nonlinear interaction, the generated signal frequency components are then combined to form the broadband output beam. Effectively, the use of the proposed scheme results in a broadband enhancement of SHG and SFG processes. We numerically show resonance-enhanced SFG exhibiting a pump conversion bandwidth of $\Delta \lambda\approx\SI{75}{nm}$ (1020--1095~nm), a value greatly exceeding the 2~nm linewidth of the SLR. 

\section{Theory} 

\label{Theory_section}

 The nonlinear response of a metasurface can be evaluated using nonlinear scattering theory~\cite{Roke2004,Obrien2015}. Using this approach, the SFG response of a metasurface depends on the mode overlap between the nonlinear polarization $\vb{P}^{(2)}(\omega_3,\vb{r})$, induced by the local fields at the fundamental frequencies $\vb{E}(\omega_1,\vb{r})$ and $\vb{E}(\omega_2,\vb{r})$, and the local field at the SFG frequency $\vb{E}(\omega_3,\vb{r})$. Consequently, the detected far-field SFG emission $\vb{E}_\mathrm{det}(\omega_3=\omega_1+\omega_2)$ can be estimated using the Lorentz reciprocity theorem as~\cite{Roke2004}
\begin{equation}
      \vb{E}_\mathrm{det}(\omega_3=\omega_1+\omega_2) \propto \iiint_V \chi^{(2)}(\omega_3;\omega_1,\omega_2,\vb{r}) : \vb{E}(\omega_1,\vb{r}) \vb{E}(\omega_2,\vb{r}) \vb{E}^*(\omega_3,\vb{r}) \,\mathrm{d}V \,,  
    \label{eq:NLST_SFG}
\end{equation}
where integration is performed over metasurface unit cell volume $V$, and $\chi^{(2)}(\omega_3;\omega_1,\omega_2,\vb{r})$ is the nonlinear susceptibility tensor. For SHG, where $\omega_1=\omega_2=\omega$ and $\omega_3=2\omega$, Eq.~\eqref{eq:NLST_SFG} is written as
\begin{equation}
      \vb{E}_\mathrm{det}(2\omega) \propto \iiint_V \chi^{(2)}(2\omega;\omega,\omega,\vb{r}): \vb{E}^2(\omega,\vb{r}) \vb{E}^*(2\omega,\vb{r}) \,\mathrm{d}V \,. 
    \label{eq:NLST_SHG}
\end{equation}
The local fields $\vb{E}(\omega_i,\vb{r})$ consist of the incident laser field $\vb{E}_\mathrm{inc}(\vb{k}_i,\omega_i)$, where $\vb{k}_i$ is the laser field wave vector, and  of the field scattered by the nanoparticles in the metasurface $\vb{E}_\mathrm{scat}(\omega_i)$. In other words, $\vb{E}(\omega_i)=\vb{E}_\mathrm{inc}(\vb{k}_i,\omega_i)+\vb{E}_\mathrm{scat}(\omega_i)$. Consequently, the local fields can be increased either by increasing the incident laser field amplitude, or by utilizing resonances, such as LSPRs or SLRs, that boost the scattered fields $\vb{E}_{\mathrm{scat}}(\omega_i)$~\cite{Huttunen2016a}.

In this work, we focus on SLRs that occur in periodic arrays of metallic NPs for two different reasons. First, SLRs can be associated with very high $Q$-factors and consequently also with considerable local-field enhancements. Second, the collective nature of SLRs makes them spatially dispersive, which can be utilized to realize broadband frequency conversion. 

Collective SLRs result from radiative coupling between periodically arranged NPs. This coupling is strong near the Rayleigh anomaly wavelength, which for the first-order diffraction mode is given by~\cite{Magnusson1998,Khlopin2017}:
\begin{equation} \label{Eq:RAangleDependence}
 \lambda_{\pm1}=P\left(n \mp\sin \theta\right)\,,
\end{equation} 
where $P$ is the array periodicity, $n$ is the refractive index of the surrounding material, and $\theta$ is the incidence angle in air, i.e., above the superstate material. Looking at Eq.~\eqref{Eq:RAangleDependence}, we see how the angle-dependence of SLRs provides simple means to tune the center wavelength of the resonance. We note that use of LSPRs does not provide similar tunability.

\begin{figure}[ht]
\includegraphics[]{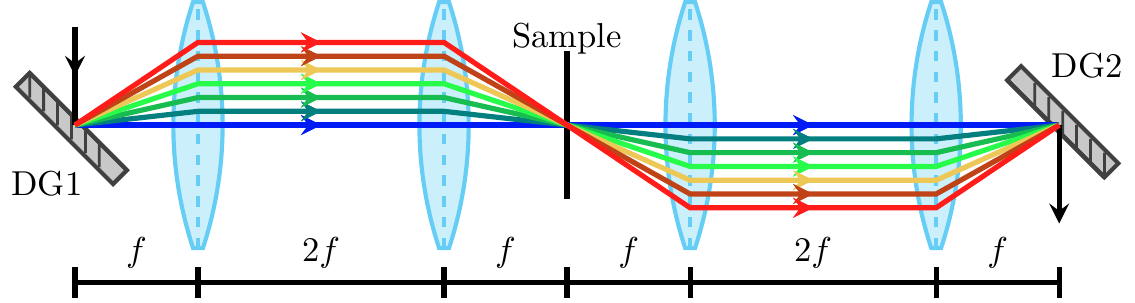}
\caption{Proposed setup for broadband frequency conversion using high-$Q$ metasurfaces. The incident pulse is separated into its spectral components using the first diffraction grating (DG1). A $4f$-correlator is then used to guide the spectral components into the metasurface, where the incidence angles $\theta$ of the spectral components depend on the frequency $\omega$. At the metasurface, the pump frequencies are up-converted with the process of sum-frequency generation. After the frequency conversion, the spectral components of the nonlinear signal are combined using a second $4f$-correlator and a subsequent diffraction grating (DG2).
}
\label{fig:Schematic}
\end{figure}

Despite the potential of utilizing high-$Q$ SLRs for enhancing light--matter interaction taking place in the metasurface, similar to all optical cavities their behavior is restricted by the time--bandwidth limit. An increase in the $Q$-factor of the cavity is necessarily associated with a reduction of the operation bandwidth. This limit seems to particularly restrict many nonlinear applications utilizing ultrashort laser pulses with pulse durations $\tau_p$ of 10--100 fs and linewidths $\Delta\lambda_L$ of 10--100 nm.
Typical high-$Q$ SLRs have linewidths $\Delta\lambda_{SLR}\,{\sim}\,1$~nm, suggesting their use with fs lasers to be inefficient. For example, when using a laser with $\tau_p\approx\SI{200}{fs}$ and $\Delta\lambda_L\approx\SI{10}{nm}$, only ${\sim}$10\,\% of the laser power can be coupled into an SLR mode with a linewidth $\Delta\lambda_{SLR}=\SI{1}{nm}$.  
To overcome this problem, we propose an experimental scheme that utilizes diffractive optical elements and the angle-dependent responses of SLRs (see Fig. \ref{fig:Schematic}). 

The scheme for broadband frequency conversion using the setup shown in Fig.~\ref{fig:Schematic} consists of five steps and resembles closely a temporal focusing scheme~\cite{Oron2005,Block2013}. First, the incident laser pulse is split into its spectral components by a diffraction grating.
The laser beam spot at the diffraction grating is then imaged using two lenses, acting as a $4f$-correlator, onto the sample plane. Thus, the different spectral components of the input beam arrive to the sample at different angles of incidence $\theta$. With a properly selected diffraction grating and a set of lenses, the incidence angle of a given frequency component can be made to match with the resonance wavelength of the tilted SLR (see Eq.~\eqref{Eq:RAangleDependence}). Therefore, it becomes possible to couple an incident broadband source more efficiently into a high-$Q$ metasurface and subsequently boost the broadband SFG response. Finally, using another pair of lenses acting as a $4f$-correlator, the SFG signal component beams are imaged onto a second diffraction grating. With a proper selection of these components, the spectral components of the SFG signal are combined with the second diffraction grating completing the broadband conversion process.

\section{Results and discussion}

\begin{figure}
    \centering
    \includegraphics{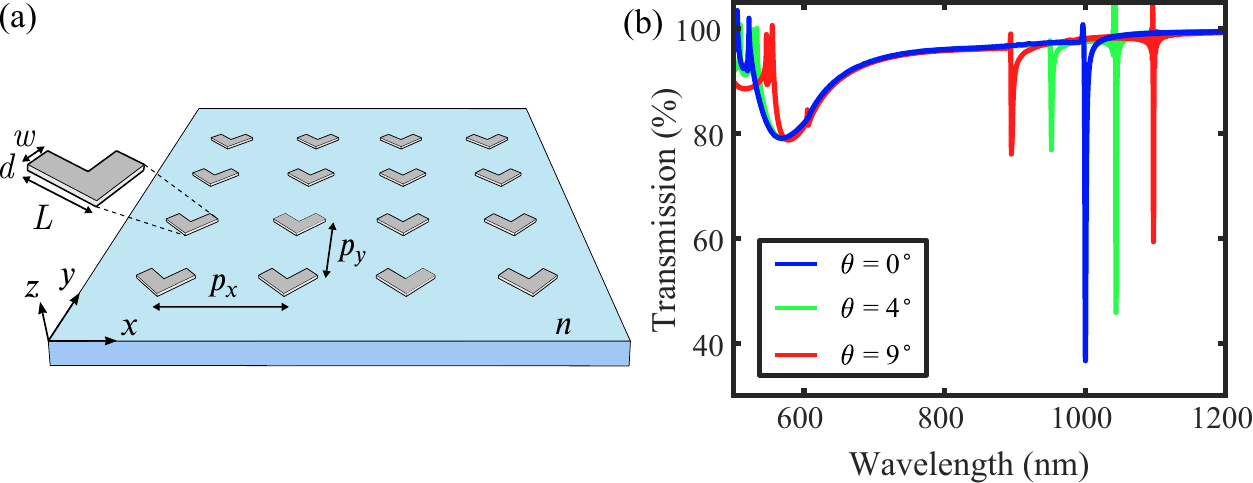}
    \caption{(a) The simulated metasurface consisting of V-shaped aluminum NPs arranged on rectangular lattice on a glass substrate ($n=1.51$). Here, the NP and lattice parameters are $L=\SI{120}{nm}$, $w=\SI{70}{nm}$, $d=\SI{30}{nm}$, $p_x=\SI{660}{nm}$, and $p_y=\SI{400}{nm}$. (b) With these parameters and $y$-polarized incident light, the metasurface exhibits a broad LSPR near 570 nm and a much narrower SLR at 1002 nm. The simulated transmission spectra shows how the peak wavelengths of the splitted SLR near 1000 nm shift from the normal incidence resonance wavelength (blue line), when the incidence angle changes. With $\theta=\ang{9}$ (red line) two SLRs form at 896 nm and 1100 nm. In this work, we chose to utilize only the SLRs occurring at longer wavelengths ($>$1000~nm), beacuse their $Q$-factors are larger than those of the shorter-wavelength SLRs ($<$1000~nm).}
    \label{fig:sample}
\end{figure}

In this work, we used finite-difference time-domain (FDTD) method to simulate the optical response of a metasurface consisting of V-shaped aluminum NPs in homogeneous glass surroundings with refractive index $n=1.51$. The NPs had arm length $L=\SI{120}{nm}$, arm width $w=\SI{70}{nm}$, and thickness $d=\SI{30}{nm}$ (see Fig.~\ref{fig:sample}\,a). The NPs were arranged in a rectangular lattice with periodicities $p_x=\SI{660}{nm}$ and $p_y=\SI{400}{nm}$ along the principal axes. For $y$-polarized light at normal incidence, our metasurface exhibited an LSPR centered at 570 nm and an SLR at 1002 nm, as is visible from the simulated transmission spectra (see Fig.~\ref{fig:sample}\,b). Here, linewidth of the SLR was $\Delta\lambda_{SLR}\approx\SI{2}{nm}$ corresponding to $Q=500$.

Next, we simulated the transmittance of the sample with the incidence angle $\theta$ varying from \ang{0} to \ang{9}. Again, we considered $y$-polarized light.
By changing $\theta$, SLR peak was split into two peaks, which moved further from 1002~nm as $\theta$ increased (see Fig.~\ref{fig:sample} b). At $\theta=\ang{9}$, SLRs occurred at 896~nm and 1100~nm. For the NP geometry considered in this work, the $Q$-factors and the local-field enhancement factors associated with the shorter-wavelength SLRs were found to be significantly lower than for those of the longer-wavelength SLRs. Therefore, in what follows, we chose to focus only on the longer-wavelength SLRs occurring between 1000--1100~nm.

\begin{figure}
    \centering
    \includegraphics{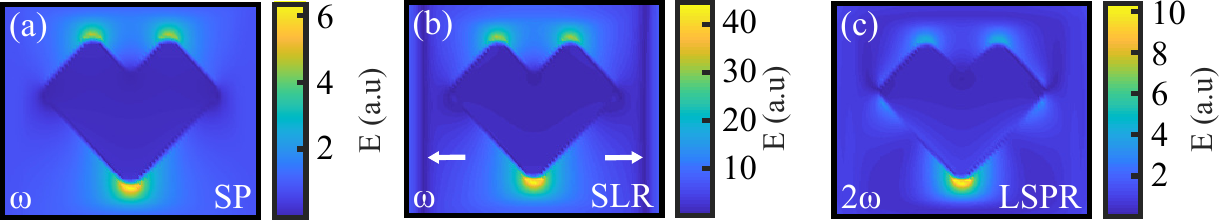}
    \caption{(a) Interaction with $y$-polarized light induces local-field hot spots on the edges of single particles (SP). (b) When nanoparticles are arranged in an array and the incident light is set to the surface lattice resonance (SLR) wavelength of 1002~nm, the local-field enhancement becomes dramatically larger than in the SP case. The collective nature of SLRs causes the field to drop symmetrically along the direction inter-particle coupling (white arrows).(c) Our nanoparticles exhibits localized surface plasmon resonances (LSPRs) in the signal wavelength range of our studies (500--550~nm). In total, the mode overlap between $E(\omega,\vb{r})$ and $E(2\omega,\vb{r})$ result in enhanced SHG and SFG responses.}
    \label{fig:local_fields}
\end{figure}

In order to verify that the local fields are accordingly enhanced in the presence of high-$Q$ SLRs, we also present the simulated local field distributions (see Fig.~\ref{fig:local_fields}). When using $y$-polarized incident light oscillating at the SLR peak wavelength, the light--matter interaction results in local-field hot spots near the corners of the NPs that point along the $y$-direction. Although the overall structure of the local-field distribution associated with a single NP (Fig.~\ref{fig:local_fields}\,a) is not markedly affected when the NPs are arranged periodically giving rise to the high-$Q$ SLR (Fig.~\ref{fig:local_fields}\,b), the field amplitudes are. Their intensity dramatically increases, when we arrange particles into an array and couple light into the SLR mode (Fig.~\ref{fig:local_fields}\,b). Clear local-field hot spots also form, when incident light is resonant with the LSPR mode (Fig.~\ref{fig:local_fields}\,c). 

Finally, we used the simulated field profiles (Fig.~\ref{fig:local_fields}) to calculate mode-overlap integrals associated with the nonlinear scattering theory (see Eqs.~\eqref{eq:NLST_SFG} and \eqref{eq:NLST_SHG}). 

\begin{figure}
    \centering
    \includegraphics[scale=1]{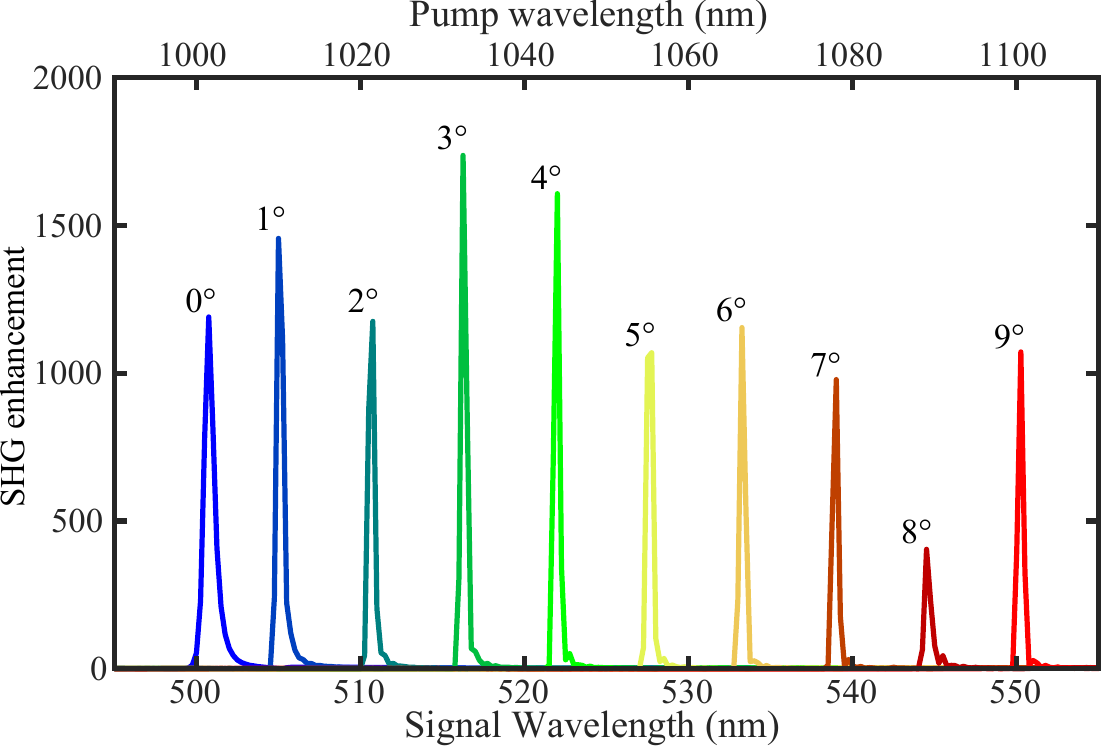}
    \caption{ Tunable frequency conversion of  narrowband laser pulses ($\Delta\lambda_L\approx\SI{1}{nm}$) via angle-dependent SHG. The SHG signals are normalized to the non-resonant case (5~nm off the SLR wavelength).}
    \label{fig:SHG_results}
\end{figure}

Here, we assumed that the nonlinear response of the plasmonic NPs is dominated by their surface response. Specifically, we only considered the susceptibility component perpendicular to the surface of the NP ($\chi^{(2)}_{\perp \perp \perp}$), and the respective field components~\cite{Wang2009}. Furthermore, the incident pump and the generated nonlinear signal fields were assumed to be polarized along the $y$-direction. 
First, we estimated how efficiently narrowband laser pulses ($\Delta\lambda_L\approx\SI{1}{nm}$, FWHM) at different wavelengths between 1000--1100 nm can be converted to SHG wavelengths between 500--550 nm (see Fig. \ref{fig:SHG_results}). Laser pulses at different wavelengths were guided on the metasurface at different incidence angles (\ang{0}--\ang{9}), resulting in enhanced SHG due to occurring SLRs. As expected by looking at Eq.~\eqref{Eq:RAangleDependence}, different incident wavelengths can be made resonant with the metasurface simply by changing the angle of incidence. 
When compared against the off-resonance situation (5~nm away from the SLR wavelength), the calculated SHG intensities were enhanced by factors in the range of 500 (\ang{8} angle of incidence) to 1800 (\ang{3} angle of incidence).

\begin{figure}
    \centering
    \includegraphics{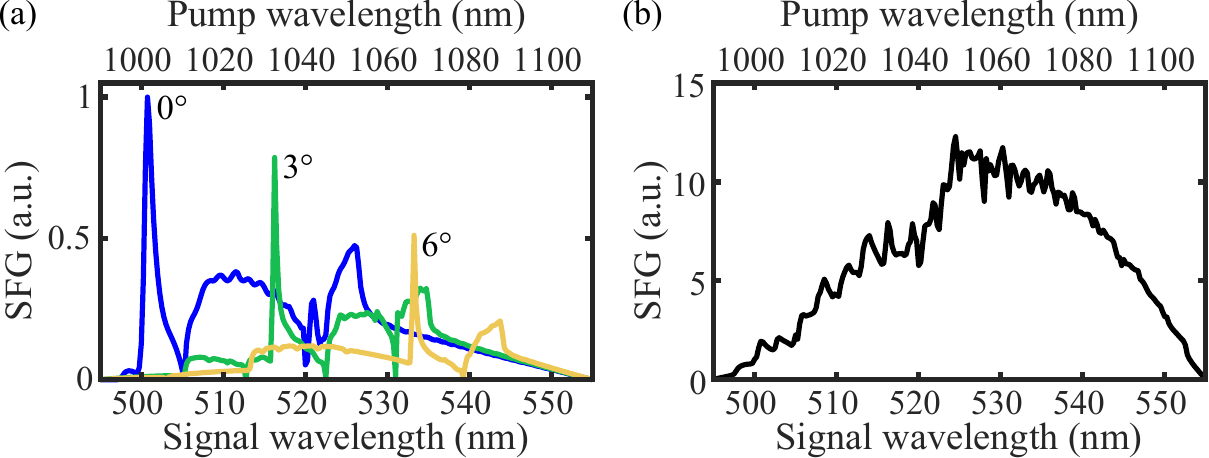}
    \caption{Frequency conversion of a broadband laser source ($\Delta\lambda_L = \SI{100}{nm}$) via SFG in an SLR-supporting metasurface ($\Delta\lambda_{SLR}\approx\SI{2}{nm}$ corresponding to $Q=500$). (a) SFG responses of studied metasurface without the proposed scheme (see Fig. \ref{fig:Schematic}). The laser pulse is incident on the metasurface at a single angle of incidence. Thus, the laser pulse arriving either at \ang{0} (blue), \ang{3} (green), or \ang{6} (yellow) will be most efficiently converted only near the corresponding SLR wavelengths 1002~nm, 1032~nm, or 1067~nm, respectively. (b) The overall SFG response of the studied metasurface using the proposed scheme. Through the SLR-induced enhancement, all spectral components interact efficiently, resulting in a broadband frequency conversion with signal bandwidth $\Delta\lambda_{SFG}\approx\SI{40}{nm}$ (pump conversion bandwidth $\Delta\lambda\approx\SI{75}{nm}$, 1020--1095~nm).}
    \label{fig:SFG_results}
\end{figure}

Second, we considered frequency conversion of a broadband laser pulse centered at wavelength of 1050 nm having a bandwidth of $\Delta\lambda_L=\SI{100}{nm}$ (FWHM). For simplicity, we assumed that the pulse is separated into 10 different spectral components each with a linewidth of $\SI{5}{nm}$. In addition, the frequency components were assumed to have constant-valued field amplitudes and to arrive at the metasurface at different incidence angles $\theta$ allowing each component to couple resonantly with an optimal SLR mode (see Fig.~\ref{fig:sample}\,b). Since the different spectral components arrive at the metasurface simultaneously, they can all interact with each other.
This gives rise to SFG at numerous different signal wavelengths for different angles of incidence (see Fig.~\ref{fig:SFG_results}\,a), at which the NLO responses are enhanced by the factor of ${\sim}1000$ when compared against the off-resonance situation.
The total SFG response of the metasurface thus becomes a sum of all individual SFG spectra (see Fig.~\ref{fig:SFG_results}\,b). It is interesting to see that due to this cumulative nature of the total SFG signal, the total SFG response is enhanced by an additional factor of $10$, when compared against calculated SFG spectra performed for individual angles of incidence. More importantly, the total SFG signal has a conversion bandwidth of $\Delta\lambda_{SFG}\approx\SI{40}{nm}$ (pump conversion bandwidth $\Delta\lambda\approx\SI{75}{nm}$, 1020--1095~nm), indicating simultaneous resonance-enhanced and broadband frequency conversion of the initial laser pulse with $\Delta\lambda_L=\SI{100}{nm}$. This pump conversion bandwidth $\Delta\lambda$ is almost 40 times broader than the linewidth $\Delta\lambda_{SLR}\approx\SI{2}{nm}$ associated with the normally incident SLR, suggesting a way to surpass the time--bandwidth limit by optimizing the way light is coupled into SLR-supporting metasurfaces.

The results above illustrate how our method results in broadband frequency conversion of light through the process of SFG. We believe that the method can be generalized to be applicable also for other nonlinear processes. For example, utilizing our approach for DFG or third-harmonic generation (THG), a broadband generation of THz or ultraviolet laser pulses could be achieved. We also note that the proposed method can be expected to be quite relevant when nonlinear responses of SLR-based metasurfaces with record-high $Q$-factors are investigated and utilized for frequency conversion applications~\cite{Saad2021}.

As a whole, this numerical work proposes a novel methodology for broadband frequency conversion of light using metasurface-based high-$Q$ resonators ($Q\approx500$). Because the time--bandwidth limit restricts the conversion bandwidth and the achievable efficiency of resonant nonlinear devices, the proposed methodology could provide new possibilities for metasurface-based broadband frequency conversion of light.

\section{Conclusions}
To conclude, we have demonstrated a method for a broadband frequency conversion using a metasurface supporting high-$Q$ SLRs ($Q\approx500$). In our proposed setup design, different wavelength components are separated and guided on an SLR-supporting metasurface at different incident angles. Due to the spatial dispersion of SLRs, the scheme results in resonance-enhanced and broadband SFG response.
Thus, our method is suitable for frequency conversion of both broadband laser pulses and of wavelength-tunable lasers with narrower spectral features.  We have shown how the frequency conversion of an ultrashort laser pulse with a linewidth of 100 nm can be resonantly enhanced (${\sim}1\,000$-fold) when comparing against non-resonant nonlinear response. Furthermore, a pump conversion bandwidth of $\Delta\lambda\approx75$~nm is achieved, exceeding by almost a factor of 40 the linewidth (2~nm) of the SLR cavity. This result suggests a way to surpass the time--bandwidth limit associated with resonant cavities by optimizing the way light is coupled into SLR-supporting metasurfaces.
In addition to SFG and SHG, our method could be generalized for other frequency conversion processes, such as DFG and THG.
Overall, our work opens new possibilities to perform broadband frequency conversion of light by utilizing high-$Q$ metasurface cavities.

\section*{Acknowledgements}
We acknowledge the support of the Academy of Finland (Grant No. 308596) and the Flagship of Photonics Research and Innovation (PREIN) funded by the Academy of Finland (Grant No. 320165). TS also acknowledges Jenny and Antti Wihuri Foundation for their PhD grant.

\section*{Data Avaibility Statement}
The data that support the findings of this work are available upon reasonable request from the authors.

\section*{Appendix A: Simulation Methods}

\textit{Linear FDTD Simulations} The FDTD simulations were performed using Lumerical FDTD Solutions simulation software. We simulated the transmission spectra and local-field distribution for aluminum nanoparticles in homogeneous surroundings ($n=1.51$). To investigate a periodic structure, we used periodic boundary conditions along the metasurface axes ($x$- and $y$-axes). The perfect-matching-layer (PML) conditions were used at the boundary along the initial propagation direction ($z$-axis).  For simulations with $\theta\neq\ang{0}$, we used the Broadband Fixed Angle Source Technique (BFAST).

\textit{Nonlinear Scattering Theory} The nonlinear responses of our metasurface was evaluated using the presented nonlinear scattering theory and Lorentz reciprocity theorem~\cite{Obrien2015,Roke2004}. The calculations were performed with Matlab. There we used the local-field profiles simulated with FDTD as input values.

\bibliography{refs}

\end{document}